\documentclass[usenatbib,useAMS]{mn2e}
\usepackage{times}
\usepackage{epsfig}
\usepackage{amsmath}
\usepackage[usenames,dvips]{color}
\usepackage{subfigure}
\usepackage{rotating}
\usepackage{graphicx}
\usepackage{lscape}
\usepackage{pdflscape}
\graphicspath{{fig/}}
\usepackage{color}

\begin{document}
\title[Polarimetric microlensing of circumstellar disks]{Polarimetric microlensing of circumstellar disks}
\author[Sedighe Sajadian and Sohrab Rahvar]
{Sedighe Sajadian$^{1}$ and Sohrab Rahvar$^{2}$ \\  \\
$^1$ School of Physics, Damghan University, P.O. Box 41167-36716, Damghan, Iran \\
$^2$ Department of Physics, Sharif University of Technology, P.O. Box 11155-9161, Tehran, Iran \\
}

\maketitle
\begin{abstract}
We study the benefits of polarimetry observations of microlensing
events to detect and characterize circumstellar disks around the
microlensed stars located at the Galactic bulge. These disks which
are unresolvable from their host stars make a net polarization
effect due to their projected elliptical shapes. Gravitational
microlensing can magnify these signals and make them be resolved.
The main aim of this work is to determine what extra information
about these disks can be extracted from polarimetry observations of
microlensing events in addition to those given by photometry ones.
Hot disks which are closer to their host stars are more likely to be
detected by microlensing, owing to more contributions in the total
flux. By considering this kind of disks, we show that although the
polarimetric efficiency for detecting disks is similar to the
photometric observation, but polarimetry observations can help to
constraint the disk geometrical parameters e.g. the disk inner
radius and the lens trajectory with respect to the disk semimajor
axis. On the other hand, the time scale of polarimetric curves of
these microlensing events generally increases while their
photometric time scale does not change. By performing a Monte Carlo
simulation, we show that almost $4$ optically-thin disks around the
Galactic bulge sources are detected (or even characterized) through
photometry (or polarimetry) observations of high-magnification
microlensing events during $10$ years monitoring of $150$ million
objects.\\
\end{abstract}


\section{Introduction}
Gravitational microlensing is assigned to the increase in the
brightness of a background star due to passing through the
gravitational field of a foreground object \cite{Einstein36}.
Although this phenomenon was first proposed as a tool to probe the
distribution and nature of dark matter in the Galactic disk
\cite{Paczynski1986}, currently microlensing has been referred to a
powerful method in detecting exo-planets, indicating the Galactic
mass distribution, probing the stellar atmospheres , etc. (see e.g.
Mao 2012, Gaudi 2012, Rahvar 2015). A significant problem in
microlensing observations is that the number of the physical
parameters of the source and lens is more than the number of the
physical quantities measured from observations (see e.g. Kains et
al. 2013). Hence, extra observational constraints are needed to
resolve the lensing degeneracy.

A method for partially breaking the microlensing degeneracy and
obtaining extra information about the lens or source star is
performing astrometry or polarimetry observations in addition to the
photometry observation. For example, the astrometric shift in the
source trajectory due to the lensing effect (see e.g. Walker 1995,
H{\o}g et al. 1995) gives us the angular Einstein radius which by
adding the parallax effect measurement, the lens mass can be
inferred \cite{Paczynski96,Miralda96,Rahvar2003}. Polarimetric
observations can also be done during microlensing events. Indeed,
there is a local polarization over the star surface due to the
photon scattering occuring in the stellar atmospheres. The effect is
particularly effective for the host stars that have a free electron
atmosphere \cite{chandrasekhar60}. By a minor extent, polarization
may be also induced in main-sequence stars (of late-type) by the
scattering of star light off atoms and molecules \cite{Fluri1999}
and in evolved, cool giant stars by photon scattering off dust
grains contained in extended envelopes \cite{simmons2002}. However,
the net polarization of a distant star is zero due to symmetric
orientations of local polarizations on its surface with respect to the source
centre. During a microlensing event, the circular symmetry of the
source surface is broken which causes a net polarization from the
source star \cite{schneider87,simmons95a,Bogdanov96}.

Polarimetry observations during microlensing events by partially
resolving the microlensing degeneracy can help us to evaluate the finite
source effect, the Einstein radius, the limb-darkening parameters of
the source star and the lens impact parameter
\cite{yoshida06,Agol96,schneider87,simmons2002}.

On the other hand, polarimetric microlensing can probe the surface
and the atmosphere of the source stars by magnify the small
polarization signals from the anomalies on the surface of stars such
as stellar spots and magnetic fields to be detected
\cite{sajadian14,sajadian2015b}. Another anomaly in the star
atmosphere that makes a net polarization is the existence of a
circumstellar disk around the source. Here, we study detecting and
characterizing this anomaly through polarimetric microlensing.

Although $170$ disks have been resolved until now
\footnote{http://www.circumstellardisks.org/}, most of them were
located at distances less than $1$ kpc from us. Through
gravitational microlensing, we can potentially detect and even
characterize the circumstellar disks around the Galactic bulge
stars. Hence, using this method we can study the environmental
effects on the disks' structure as well as statistically investigate
them. Gravitational microlensing of (or due to) disks has been
discussed in details by a number of authors, e.g. microlensing by
lenses consisting of the gas clouds was first investigated by Bozza
and Mancini (2002). Then, microlensing effects on a source
surrounded by a circumstellar disk as a function of the wave length
were studied by Zheng and M\'enard (2005) and they showed that the
magnification factors of these disks reach to $\mathrm{10\%-20\%}$
in the mid- and far- infrared. Recently, Hundertmak et al. (2009)
investigated the microlensing as a tool for detecting debris disks
around microlenses and estimated that one debris disk per year can
be detected through microlensing. Another application of
gravitational microlensing for detecting circumstellar disks around
source stars is performing polarimetry observations.

In this work, we investigate detecting and characterizing
circumstellar disks around source stars \emph{through polarimetry
observations} during microlensing events. Here, only optically-thin
disks which contain hot circumstellar dusts and are located beyond
the condensation radius $\mathrm{(\sim 1500 K)}$ are considered. It
seems that about $10\%$ of solar-type stars (G- and K-type) in our
neighbourhood have such disks \cite{Absil2013,Ertel2014}. Detecting
this kind of disks even around nearby stars needs high-resolution
observations with interferometers, because of small angular
separation between the disk and the host star. In the direction of
Galactic bulge for polarimetry microlensing, the late-type stars
with disks around them are the most suitable sources for polarimetry
observations. Our plan is also performing statistical study of these
events. During the lensing of these sources, the closer distances
from their host stars (small impact parameter) makes higher
magnification factor as well as higher contributions in total Stokes
parameters.

In section (\ref{form}) we explain the formalism used to calculate
the polarization of sources surrounded by circumstellar disks in
microlensing events. The characteristics of the polarimetric
microlensing of source stars with circumstellar disks are studied in
section (\ref{two}). In section (\ref{Monte}) we evaluate the
efficiencies of detecting disks around the source stars through
polarimetry and photometry microlensing (in optical wavelength) by
performing a Monte Carlo simulation. Finally, we give conclusions in
the last section.

\section{The formalism of polarimetric microlensing of
disks}\label{form} Here, we explain how to calculate the
polarization in gravitational microlensing (the first subsection)
and the polarization due to lensed source stars with circumstellar
disks (the second subsection).

\subsection{Polarimetry microlensing}
The existence of a net polarization due to the lensing effect for
supernovae was first pointed out by Schneider \& Wagoner (1987).
They analytically estimated the amounts of polarization degree near
point and critical line singularities. The polarization during
microlensing events was investigated by a number of authors
\cite{simmons95a,Bogdanov96}. Polarization in binary microlensing
events was also numerically calculated by Agol (1996) and he noticed
that in a binary microlensing event the net polarization is larger
than the net polarization generated by a single lens and can reach
to one per cent during the caustic-crossing. Polarization from
microlensing of cool giants with spherically symmetric envelopes by
a point lens and binary microlenses was investigated by Simmons et
al. (2002) and Ignace et al. (2006). Ingrosso et al. (2012,2015)
evaluated the expected polarization signals for a set of reported
high-magnification single-lens and exo-planetary microlensing events
as well as OGLE-III events towards the Galactic bulge. Recently,
Sajadian and Rahvar (2015) noticed that there is an orthogonal
relation between the polarization and astrometric shift of source
star position in the simple and binary microlensing events except in
the fold singularities and investigated the advantages of this
correlation for studying the surface of a source star and spots on
it.

To describe a polarized light, we use the Stokes parameters
$\mathrm{S_{I}}$, $\mathrm{S_{Q}}$, $\mathrm{S_{U}}$ and
$\mathrm{S_{V}}$. These parameters are, the total intensity, two
components of linear polarizations and circular polarization over
the source surface, respectively \cite{Tinbergen96}. Taking into account that there
is only linear polarization of light scattered on the atmosphere of a star, we set
$\mathrm{S_{V}=0}$. Therefore, the polarization degree $(P)$ and
angle of polarization $\mathrm{(\theta_{p})}$ as functions of total
Stokes parameters are \cite{chandrasekhar60}:
\begin{eqnarray}\label{eq2}
P&=&\frac{\sqrt{S_{Q}^{2}+S_{U}^{2}}}{S_{I}},\nonumber\\
\theta_{p}&=&\frac{1}{2}\tan^{-1}{\frac{S_{U}}{S_{Q}}}.
\end{eqnarray}
In microlensing events, the Stokes parameters are given by:
\begin{eqnarray}\label{tsparam}
S_{I}&=&~\rho^2_{\star}\int_{0}^{1}\rho~d\rho\int_{-\pi}^{\pi}d\phi I^{\star}(\mu)~ A(u),\\
\left( \begin{array}{c} S_{Q}\\
S_{U}\end{array}\right)&=&\rho^2_{\star}\int_{0}^1\rho~d\rho\int_{-\pi}^{\pi}d\phi I^{\star}_{-}(\mu) A(u) \left( \begin{array}{c} -\cos 2\phi \nonumber \\
\sin 2\phi \end{array} \right),
\end{eqnarray}
where $\mathrm{\rho}$ is the distance from the center to each
projected element over the source surface normalized to the
projected radius of star on the lens plane (i.e.
$\mathrm{\rho_{\star}}$), $\mathrm{\mu=\sqrt{1- \rho^{2}}}$,
$\mathrm{\phi}$ is the azimuthal angle between the lens-source
connection line and the line from the center to each element over
the source surface, $\mathrm{u=(u_{cm}^2+ \rho^2 \rho^{2}_{\star}-2
\rho \rho_{\star} u_{cm} \cos\phi)^{1/2}}$ is the distance of each
projected element over the source surface with respect to the lens
position, $\mathrm{u_{cm}}$ is the impact parameter of the source
center and $A(u)$ is the magnification factor.

$\mathrm{I^{\star}(\mu)}$ and  $\mathrm{I^{\star}_{-}(\mu)}$ are the
total and polarized light intensities that depends on the type of
the source star. Different physical mechanisms for various types of
stars make the overall polarization signals \cite{Ingrosso12}. For
instance, in the hot early-type stars, the electron scattering in
their atmosphere produces the polarization signal. The amounts of
Stokes intensities over the surface of these stars were first
evaluated by Chandrasekhar (1960) and then their numerical amounts
were approximated by the following functions \cite{schneider87}:
\begin{eqnarray}\label{II}
I^{\star}(\mu)&=&I_{0}(1-c_{1}(1-\mu)),\nonumber\\
I^{\star}_{-}(\mu)&=&I_{0}c_{2}(1-\mu),
\end{eqnarray}
where $\mathrm{c_{1}=0.64}$, $\mathrm{c_{2}=0.032}$ and
$\mathrm{I_{0}}$ is the source intensity in the line of sight
direction. For late-type main-sequence stars, the polarization
signal is produced because of both Rayleigh scattering on neutral
hydrogens and with a minor contribution from the Thompson scattering by free
electrons \cite{Fluri1999}. The amount of polarization degree for
this kind of stars was approximated as follows \cite{Stenflo2005}:
\begin{eqnarray}
P(\mu)=q_{\lambda}\frac{1-\mu^{2}}{(\mu+m_{\lambda})(I_{\lambda}(\mu)/I_{\lambda}(1))},
\end{eqnarray}
where $\mathrm{I_{\lambda}(\mu)/I_{\lambda}(1)}$ shows the
center-to-limb variation of the intensity, $\mathrm{q_{\lambda}}$
and $\mathrm{m_{\lambda}}$ are linear functions of wavelength (see
more details in Ingrosso et al. 2012). Finally, in cool giant stars
Rayleigh scattering on atomic and molecular species or on dust
grains generates the polarization signal. In this regard, Simmons et
al. (2002) offered the relevant Stokes parameters for giant stars
with spherically circumstellar envelopes lensed by a single lens.

\subsection{Polarization due to circumstellar disks}
We assume that there is a circumstellar disk around the source star.
A disk has mostly a circular shape with respect to its host star,
while its projected shape on the sky plane is generally an ellipse.
Hence, the projection breaks the circular symmetry and the result is
a net polarization. Gravitational lensing of the star surrounded by
such a disk can magnify the disk polarization signal, change its
orientation and make it be detected. The amount of magnified
polarization depends on the orientation of disk with respect to the
lens trajectory as well as the disk structure and its projected
surface density.

Here we use some parameters to quantify a circumstellar disk around
the source star in our calculation as: (i) the inner and outer radii
$\mathrm{R_{i}}$, $\mathrm{R_{o}}$ (here, we define the disk inner
radius beyond the source condensation radius where the dust forms
and the outer radius of the disk where its Stokes intensities with
respect to the source intensity decrease significantly e.g. by four
orders of magnitude), (ii) the inclination angle $i$ to project the
disk on the sky plane, i.e. the angle between the normal to disk and
the line of sight towards the observer and (iii) some parameters to
indicate the disk density distribution: the length scale
$\mathrm{R_{c}}$ where the electron number density of the disk in
the mid plane reaches to $\mathrm{n_{0}}$ (here we set
$\mathrm{R_{c}=R_{i}}$), $\mathrm{\gamma}$ and $\mathrm{\beta}$
which specify the radial dependence of the disk surface density and
its thickness respectively, $\mathrm{R_{s}}$ and $\mathrm{h_{0}}$
which indicate the thickness of the disk versus the radial distance
in the mid plane. The more details about the disk density
distribution can be found in the Appendix (\ref{density}).

We also assume some limitations to model the Stokes intensities of
the disk: (i) the disk has a circular shape in the stellar
coordinate system, (ii) the disk is optically thin, (iii) The
scattering opacity which produces the polarization for the disk is
the photon scattering on atomic and molecular species (Rayleigh
scattering) or on the dust grains, (iv) the magnetic field of the
disk is negligible, (v) the incident radiation of the source star to
the disk is unpolarized, (vi) any diffuse contribution from other
parts of the disk is not included, (vii) the single scattering
approximation is used and (viii) the disk inner radius is beyond of
the condensation radius.

Let us characterize the lens plane by $\mathrm{(x,y)}$ axes and put
the projected center of source at the center of coordinate system so
that $\mathrm{x}$-axis is parallel with the semimajor axis of the
disk (after projection on the sky and lens planes),
$\mathrm{y}$-axis is normal to it and $\mathrm{z}$-axis is toward
the observer (see Figure \ref{fig3}). We normalize all parameters to
the Einstein radius of the lens, $\mathrm{R_{E}}$. Generally, the
projected disk on the sky plane has an elliptical shape that the
ratio of its minor to major axes is $\mathrm{\cos(i)}$. The
positions of the lens and each element of the projected disk in this
reference frame normalized to $\mathrm{R_{E}}$ are
$\mathrm{(x_{l},y_{l})}$ and $\mathrm{(x_{d},y_{d})=(\varrho \cos
\varphi,\varrho \sin\varphi)}$ where $\mathrm{\varphi}$ alters in
the range of $\mathrm{[0,2\pi]}$ and $\mathrm{\varrho}$ changes in
the range of $\mathrm{\varrho \in [\varrho_{i},\varrho_{o}](1+
\sin^{2}\varphi\tan^{2}i)^{-1/2}}$ \cite{zheng2005},
$\mathrm{\varrho_{i}=R_{i}x_{ls}/R_{E}}$ and
$\mathrm{\varrho_{o}=R_{o}x_{ls}/R_{E}}$ where
$\mathrm{x_{ls}=D_{l}/D_{s}}$ the ratio of the lens and the source
distances from the observer. Indeed, we choose a circular disk
around the source star before projecting on the sky plane.

To calculate the overall Stokes parameters, we should add the
contributions of the source and its disk as follows:
\begin{eqnarray}\label{main}
\left(\begin{array}{c}S'_{I}\\S'_{Q}\\S'_{U}\end{array}\right)&=&
\left(\begin{array}{c}S_{I}\\S_{Q}\\S_{U}\end{array}\right)\nonumber\\&+&
\int_{0}^{2\pi}~d\varphi~\int_{\varrho_{i}}^{\varrho_{o}}~\varrho~d\varrho~A(u_{d})\boldsymbol{I}_{d}(\varrho,\varphi),
\end{eqnarray}
where $\mathrm{S_{I}}$, $\mathrm{S_{Q}}$ and $\mathrm{S_{U}}$ are
the Stokes parameters due to the source given by equation
(\ref{tsparam}) and the second term calculates the disk Stokes
parameters. $\mathrm{u_{d}=\sqrt{(x_{d}-x_{l})^2+(y_{d}-y_{l})^2}}$
is the distance of each projected element of the disk from the lens
position. Integrating is done over the disk area projected on the
lens plane. $\mathrm{\boldsymbol{I}_{d}(\varrho,\varphi)}$ as a vector contains the
Stokes intensities of the disk in the observer coordinate system.
More details about calculating these Stokes intensities by
considering the mentioned limitations are brought in the Appendix
(\ref{appen1}).

\begin{figure}
\begin{center}
\psfig{file=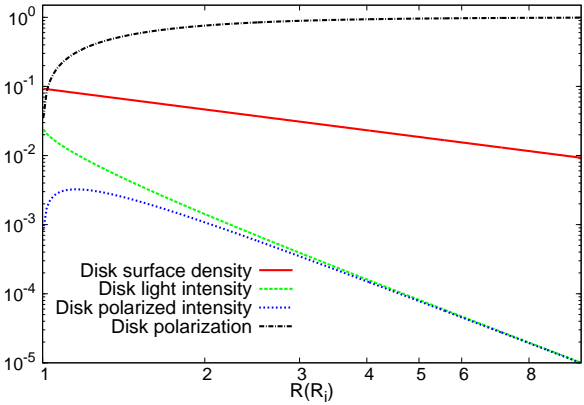,angle=270,width=8.cm,clip=} \caption{The
surface density (normalized to the amount of $n_{0}$) (red solid
line), light intensity (green dashed line), polarized intensity
(blue dotted line) and polarization signal (black dot-dashed line)
of the disk are plotted versus the normalized distance started from
the disk inner radius. Here, we assume that the disk and central
star intensities are resolvable, so that we mask the central star
and consider only the light and polarized intensities from the disk
to calculate the polarization degree. Here, we set
$\mathrm{\gamma=9/4}$, $\mathrm{\beta=5/4}$, $\mathrm{h_{0}=10 AU}$,
$\mathrm{R_{c}=R_{i}=R_{\star}}$, $\mathrm{R_{s}=100 AU}$,
$\mathrm{I_{0}=1}$ and $\mathrm{\tau_{eq}=1}$.} \label{fig1}
\end{center}
\end{figure}
\begin{figure}
\begin{center}
\psfig{file=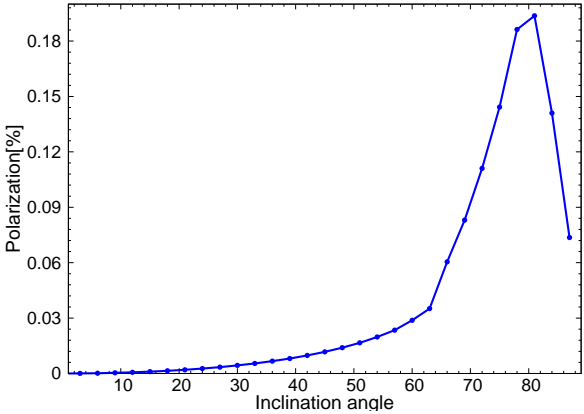,angle=270,width=8.cm,clip=} \caption{The total
polarization signal owing to a source star with the circumstellar
disk characterized in Figure (1) without lensing effect versus the
disk inclination angle.} \label{fig2}
\end{center}
\end{figure}

Here, we study the disk polarization signal without considering the
lensing effect. In Figure (\ref{fig1}) we plot the surface density
(red solid line), light intensity (green dashed line), polarized
intensity (blue dotted line) and polarization degree (black
dot-dashed line) of a typical disk versus the normalized distance
started from the inner radius of disk. The parameters used to make
this figure are $\mathrm{\gamma=9/4}$, $\mathrm{\beta=5/4}$,
$\mathrm{h_{0}=10 AU}$, $\mathrm{R_{s}=100 AU}$, $\mathrm{I_{0}=1}$,
$\mathrm{R_{c}=R_{i}=R_{\star}}$ where $\mathrm{R_{\star}}$ is the
source radius. Note that we normalize the disk surface density to
the amount of $\mathrm{n_{0}}$ (whose definition is brought in
equation \ref{n0}) to justify its range with the ranges of the light
and polarized intensities. As explained in the Appendix
(\ref{density}), the polarization signal due to the disk is
proportional to the equatorial optical depth due to the dust
scattering, i.e. $\mathrm{\tau_{eq}}$, whose definition is brought
in the equation (\ref{etau}). Here we set $\mathrm{\tau_{eq}=1}$.
Here, we assume that the disk and the central star intensities are
resolvable, so that we mask the central star and consider only the
light and polarized intensities from the disk for calculating the
polarization degree. The light intensities as well as the disk
surface density decrease by increasing the distance from the source
star. The disk polarized light is maximized at almost the disk inner
radius. Hence, during lensing if the lens crosses the disk inner
radius the polarimetric disk-induced perturbation maximizes.

If the source is so far from us, we can not discern the source light
from the disk light and receive the total polarization signal due to
the disk and its central star. If there is no lensing effect, the
total polarization signal depends strongly on the projected shape of
the disk which is a function of the disk inclination angle as well
as the dust optical depth of the disk. In Figure (\ref{fig2}) we
plot the total polarization signal of the source star surrounded by
the circumstellar disk (specified in Figure \ref{fig1}) versus the
disk inclination angle. The more inclined disks, the higher
polarization signals. However, the maximum polarization signal
happens for the disks with the inclination angles between
$\mathrm{70^{\circ}-80^{\circ}}$ which agrees with the results of
simulations of polarization signals for gaseous disks around
Be-stars (see e.g. Halonen et al. 2013, Halonen and Jones 2013). The
polarization angle of different inclined disks does not depend on
the disk inclination angle in contrast with the polarization degree.
Because, an inclined disk has a symmetric density distribution with
respect to its semiminor and major axes. Hence, the disk Stokes
parameter $\mathrm{S_{U}}$ vanishes and $\mathrm{S_{Q}}$ becomes
negative which result the polarization angle equal to $90$ with
respect to its semimajor axis (see equation \ref{eq2}).
\begin{figure}
\begin{center}
\psfig{file=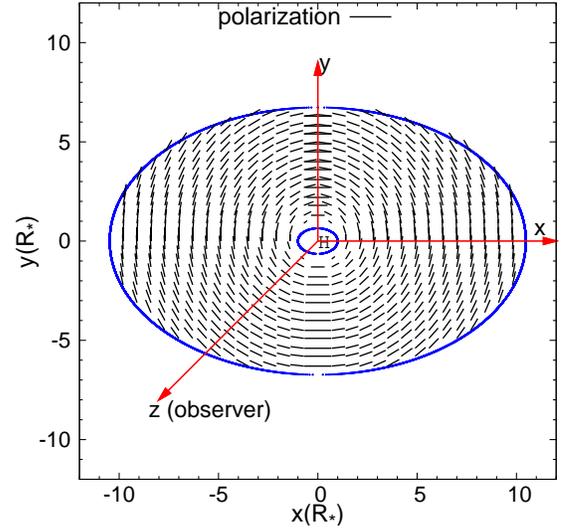,angle=270,width=10.cm,clip=} \caption{The
polarization map around that disk specified in Figure (\ref{fig1})
with the inclination angle $\mathrm{i=50^{\circ}}$. The size of
lines is proportional to the polarization degree.} \label{fig3}
\end{center}
\end{figure}

In Figure (\ref{fig3}) we plot the polarization map over that disk
with the inclination angle $\mathrm{i=50^{\circ}}$. Note that, the
size of lines is proportional to the polarization degree. The
polarization orientation (i.e. $\mathrm{\theta_{p}}$) is given in
terms of its angle with respect to $\mathrm{x}$-axis. The largest
polarization signals occur near the disk outer radius whereas the
maximum polarized intensity comes from the disk inner radius. Having
simulated a circumstellar disk around the source star, we turn on
the lensing effect and study detecting disks through polarimetric
microlensing which is explained in the following section.

\section{Polarimetric microlensing of disks}\label{two}
\begin{figure*}
\begin{center}
\subfigure[] {
\psfig{file=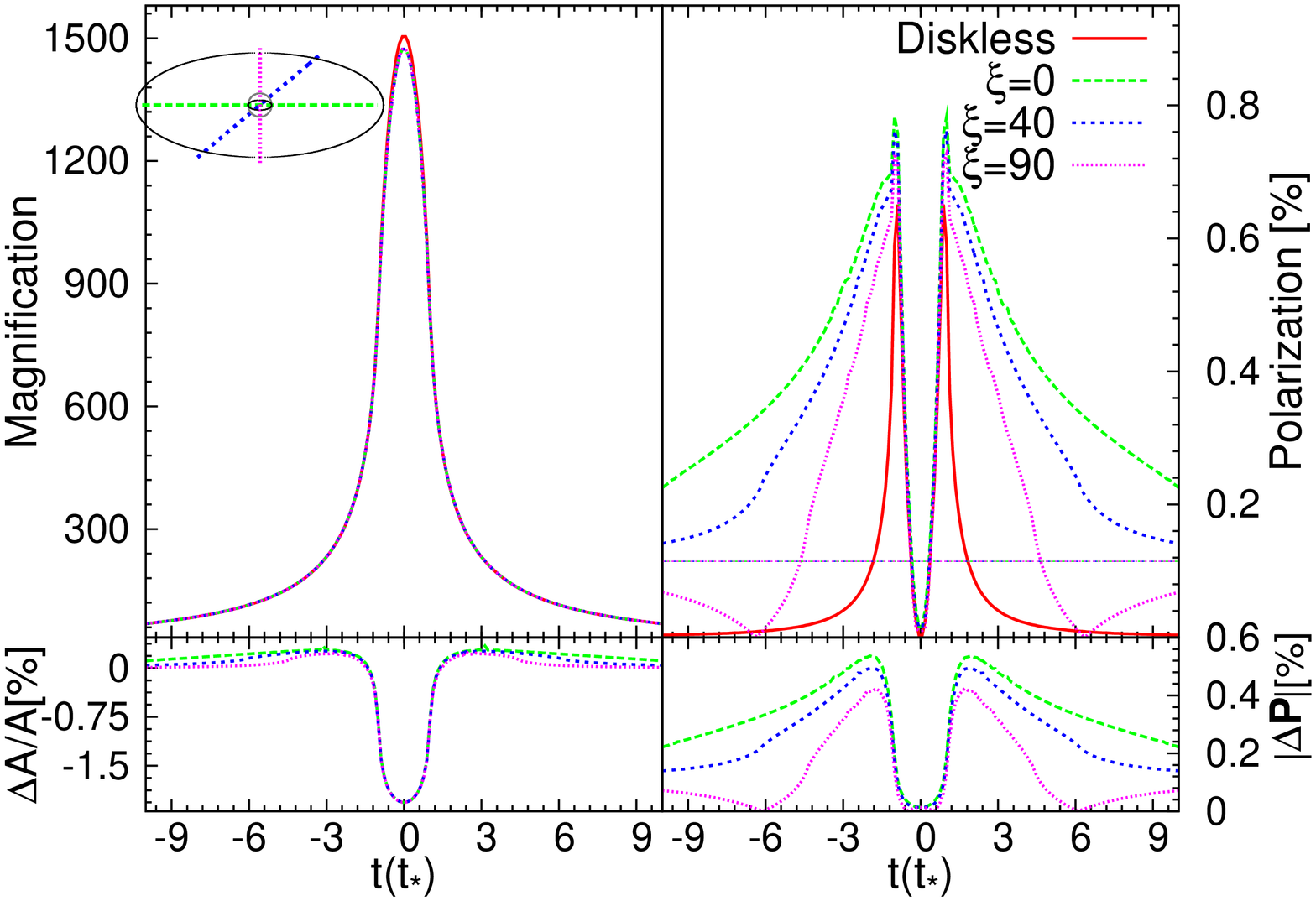,angle=270,width=0.45\textwidth,clip=0}\label{fig4a}
} \subfigure[] {
\psfig{file=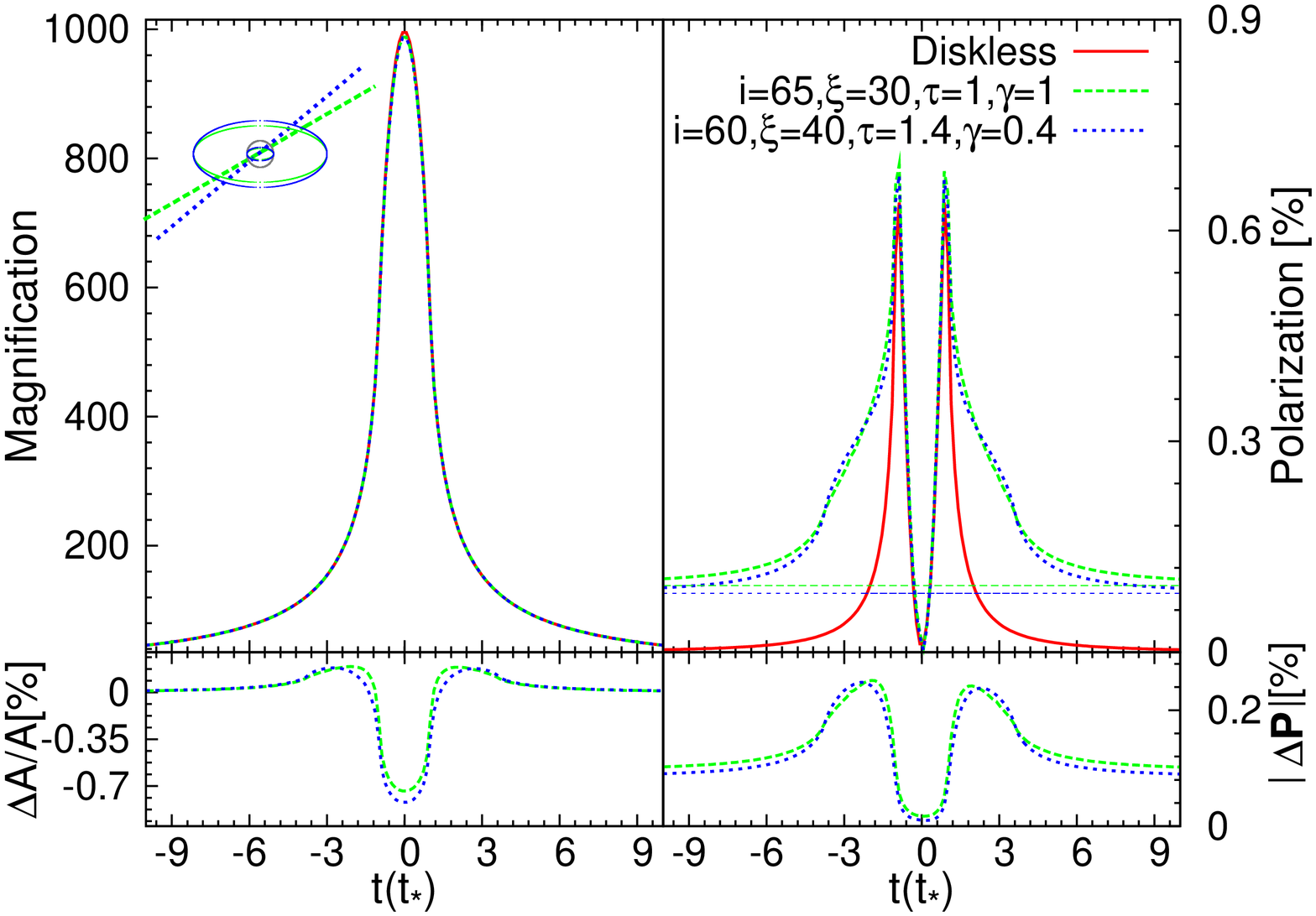,angle=270,width=0.45\textwidth,clip=0}\label{fig4b}}
\subfigure[] {
\psfig{file=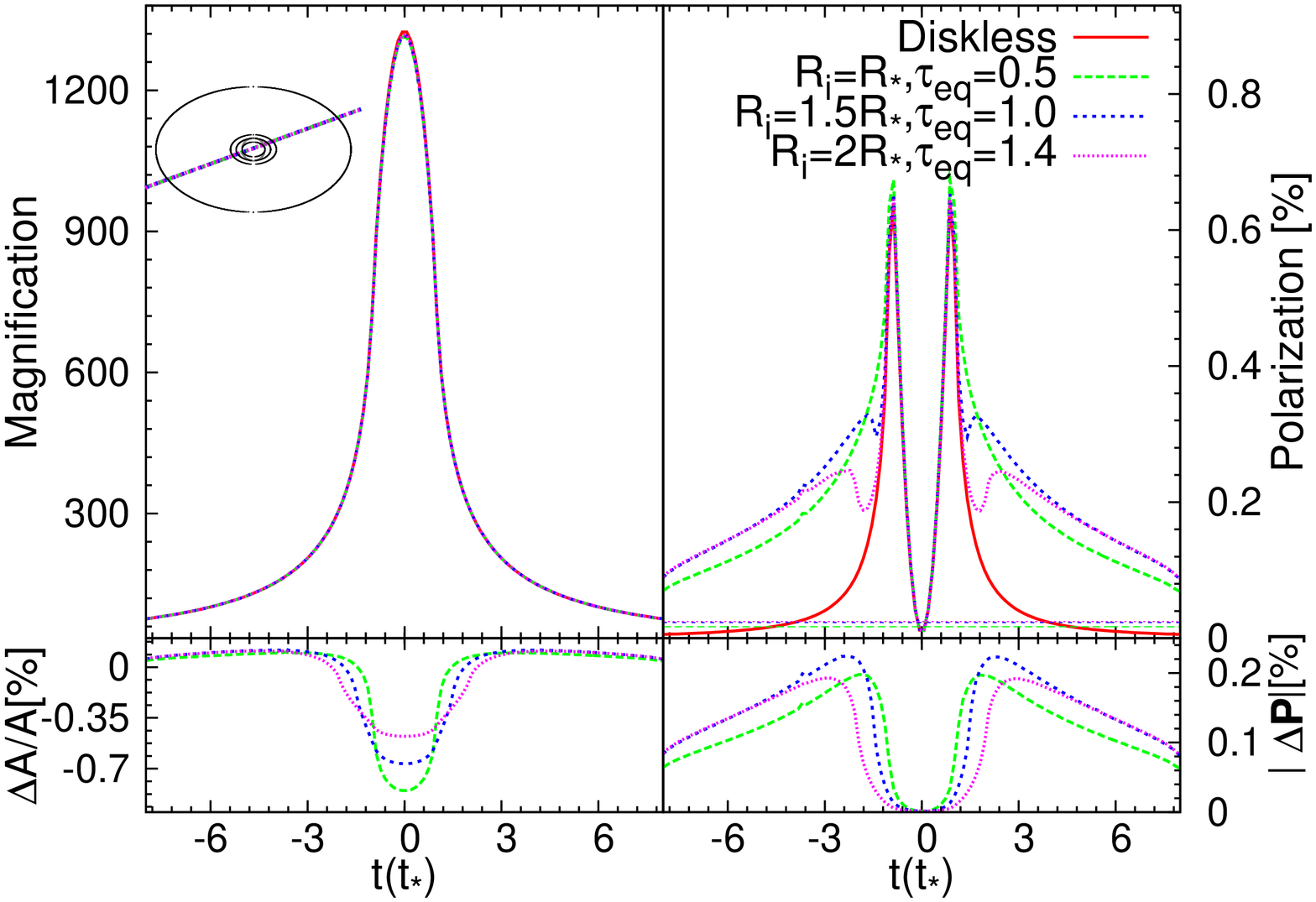,angle=270,width=0.45\textwidth,clip=0}\label{fig4c}
} \subfigure[] {
\psfig{file=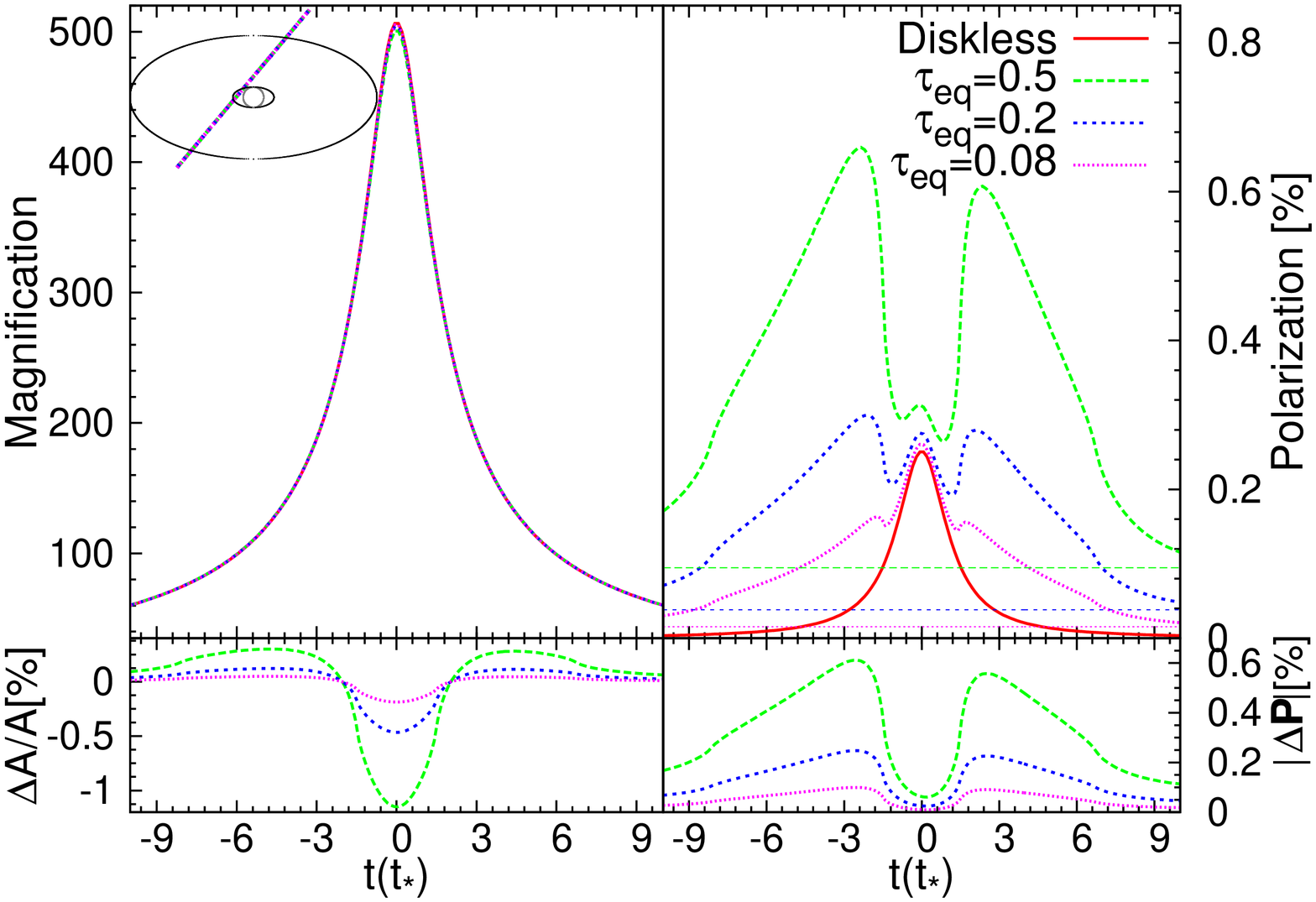,angle=270,width=0.45\textwidth,clip=0}\label{fig4d}
} \subfigure[] {
\psfig{file=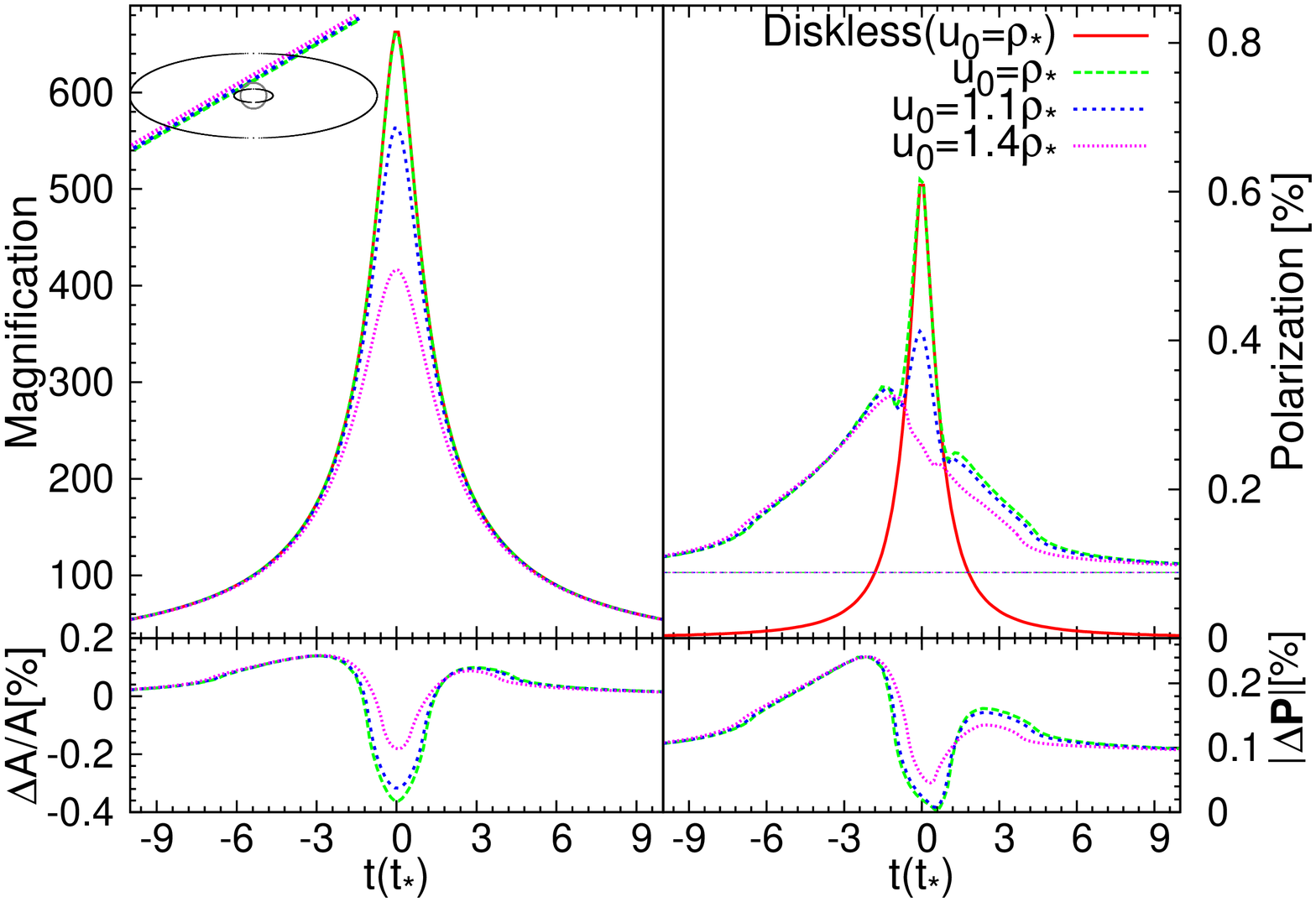,angle=270,width=0.45\textwidth,clip=0}\label{fig4e}
} \subfigure[] {
\psfig{file=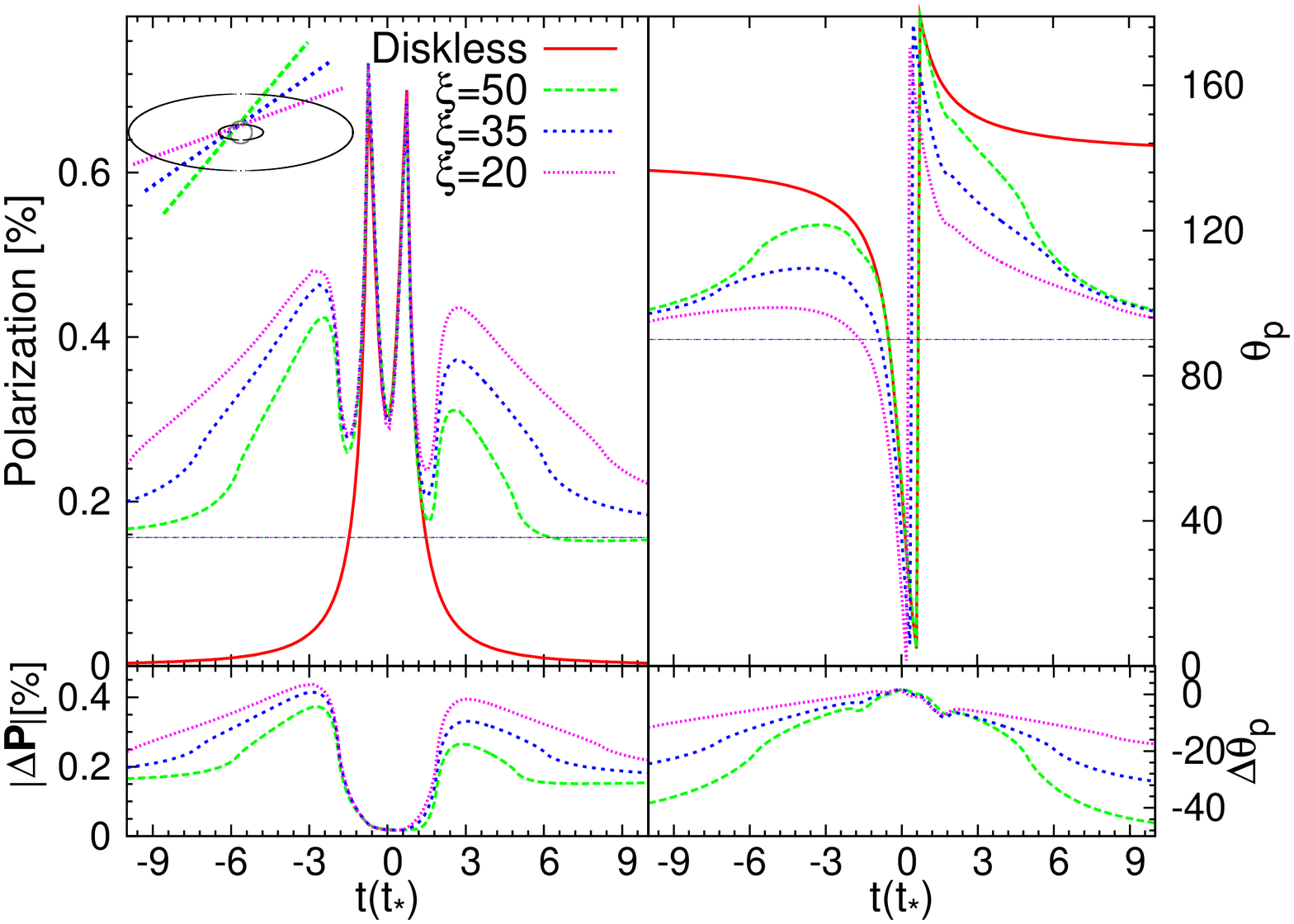,angle=270,width=0.45\textwidth,clip=0}\label{fig4f}}
\caption{Example polarimetric microlensing events affected with a
circumstellar disk around the source star. In each subfigure, the
light and polarimetric curves are shown in left and right panels.
The source (grey circle) with circumstellar disk (black ellipses)
projected on the lens plane and lens trajectory are shown with
insets in the left-hand panels. The simple models without disk
effect are shown by red solid lines. The thinner straight lines in
right panels represent the polarization signal of disk without
lensing effect. The photometric and polarimetric residuals with
respect to the simple models are plotted in bottom panels. Noting
that, in the last panel (f), the curves of polarization degree and
its angle during a microlensing event with three different amounts
of $\mathrm{\xi}$ are shown instead of light and polarimetric
curves. The parameters used to make these figures can be found in
Table (\ref{tab1}). We also set $\mathrm{M_{l}=0.3~M_{\odot}}$,
$\mathrm{D_{l}=6.5~kpc}$, $\mathrm{D_{s}=8~kpc}$,
$\mathrm{R_{c}=R_{i}}$, $\mathrm{h_{0}=10~AU}$,
$\mathrm{R_{s}=100~AU}$ and the limb-darkening coefficients over the
source surface $\mathrm{c_{1}=0.64}$ and
$\mathrm{c_{2}=0.032}$.}\label{fig4}
\end{center}
\end{figure*}

In this section, we study the characteristics of polarimetric curves
of lensed source stars with circumstellar disks. In Figure
(\ref{fig4}) some polarimetry and photometry curves due to
microlensing events of source stars surrounded by circumstellar
disks are represented. According to these Figures, we notice some
significant properties of light and polarimetric curves as
following. In this section, we consider early-type stars in the
microlensing sources and use equations (\ref{II}) to calculate the
source Stokes intensities.

(a) The time scale of the polarimetric curve of the lensed source
star surrounded by a circumstellar disk increases. This time scale
depends on (i) the disk density distribution which indicates the
disk equatorial optical depth and (ii) the disk and lens geometrical
parameters, e.g. the angle of the lens-source trajectory with
respect to the disk semimajor axis $\mathrm{\xi}$ and the disk
inclination angle. The disk length that lens crosses with the impact
parameter $\mathrm{u_{0}}$ is given by:
\begin{eqnarray}\label{len}
\Delta L=\frac{2}{f^2}\sqrt{\varrho^2_{o}f^2-u^2_{0}(1+\tan^2 i)},
\end{eqnarray}
where $\mathrm{f=\sqrt{1+\sin^{2}\xi \tan^{2}i}}$. Examples of these
lengths are shown in the inset in the left-hand panel of Figure
\ref{fig4a} in which the ellipse shows the projected outer radius of
the disk and the straight lines represent the lens trajectories with
the different angles with respect to the disk semimajor axis. Let us
assume $\mathrm{\varrho_{o}\sim 5 \rho_{\star}}$ the distance from
the source star in which the polarized intensity of the disk with
$\mathrm{\tau_{eq}=1}$ almost decreases by four orders of magnitude
with respect to the source intensity (according to Figure
\ref{fig1}). By considering high-magnification microlensing events
with $\mathrm{u_{0}\leq \rho_{\star}}$ in which the probability of
detecting the polarization signals due to the lensed sources is
high, we calculate the average amount of the time duration when the
lens crosses the disk length which is $\mathrm{\bar{\Delta
t}=\bar{\Delta L}\times t_{E}\sim 8 t_{\star}}$. Here
$\mathrm{t_{E}}$ is the Einstein crossing time and
$\mathrm{t_{\star}=\rho_{\star} t_{E}}$ is the time of crossing the
projected source radius which is the time scale of a typical
polarimetric microlensing without any perturbation. Noting that
$\mathrm{\bar{\Delta L}}$ is a dimensionless parameter. Therefore,
the time scale of the polarimetric microlensing of the source
surrounded by the disk with $\mathrm{\tau_{eq}=1}$ increases in
averagely by one order of magnitude with respect to that without
disk effect. If the disk equatorial optical depth decreases, the
outer radius of the disk shrinks. As a result, the increase in the
polarimetric time scales lessens by decreasing the disk equatorial
optical depth. Increasing in the time scale also occurs in the
polarimetric microlensing events of cool giant stars with
spherically symmetric envelopes \cite{simmons2002}. We note that the
time scale of light curves of these microlensing events does not
significantly change in contrast with the polarimetric ones.

\begin{table*}
\begin{center}
\begin{tabular}{|c|c|c|c|c|c|c|c|c|c|c|c|c|c|}
& Figure number & $M_{\star}(M_{\odot})$ & $R_{\star}(R_{\odot})$ &
$T_{\star}(K)$ & $\rho_{\star}\times 10^{3}$ & $R_{i}(R_{\star})$ &
$R_{o}(R_{\star})$ & $\tau_{eq}$ & $i^{\circ}$ & $\gamma$& $u_{0}(\rho_{\star})$  &$\xi^{\circ}$ &\\
\hline\hline
& $4(a)$ & $0.6$ & $0.66$ & $3796$  & $1.5$ & $1$ & $10.5$ & $1$ & $65$ & $1.0$ & $0$ & $--$ &\\
\hline
& $4(b)$ & $1.0$ & $1.0$ & $5770$ & $2.2$ & $1$ & $5.5$ & $--$ & $--$ & $--$ & $0.1$ & $--$ &\\
\hline
& $4(c)$ & $0.7$ & $0.75$ & $4191$  & $1.6$ & $--$ & $8.5$ & $--$ & $50$ & $1$ & $0.1$ & $20$ &\\
\hline
& $4(d)$ & $0.7$ & $0.75$ & $4191$  & $1.6$ & $2$ & $10.5$ & $--$ & $60$ & $1$ & $1.3$ & $50$ & \\
\hline
& $4(e)$ & $0.8$ & $0.83$ & $4719$ & $1.8$ & $1.5$ & $9.5$ & $1.1$ & $70$ & $1$ & $--$ & $30$ & \\
\hline
& $4(f)$ & $0.7$ & $0.75$ & $4191$ & $1.6$ & $2$ & $10.5$ & $0.3$ & $70$ & $1$ & $0.6$ & $--$ & \\
\hline
\end{tabular}
\end{center}
\caption{The table contains the parameters used to make microlensing
events shown in Figures \ref{fig4a}, \ref{fig4b}, \ref{fig4c},
\ref{fig4d}, \ref{fig4e} and \ref{fig4f}. These parameters are the
source mass $\mathrm{M_{\star}(M_{\odot})}$, the source radius
$\mathrm{R_{\star}(R_{\odot})}$, the photosphere temperature of the
source $\mathrm{T_{\star}(K)}$, the projected source radius in the
lens plane, normalized to the Einstein radius
$\mathrm{\rho_{\star}(\times 10^{3})}$, the disk inner radius
normalized to the source radius $\mathrm{R_{i}(R_{\star})}$, the
normalized outer radius of the disk $\mathrm{R_{o}(R_{\star})}$, the
disk equatorial optical depth $\mathrm{\tau_{eq}}$, the inclination
angle of the disk $\mathrm{i^{\circ}}$, $\mathrm{\gamma}$ which
indicates the radial dependence of disk density distribution, the
impact parameter of the source center normalized to
$\mathrm{\rho_{\star}}$ $\mathrm{u_{0}(\rho_{\star})}$ and the angle
of the lens trajectory with respect to the disk semimajor axis
$\mathrm{\xi^{\circ}}$ respectively. For these figures, we also set
the mass of the primary lens $\mathrm{M_{l}=0.3~M_{\odot}}$, the
lens and source distance from the observer $\mathrm{D_{l}=6.5~kpc}$
and $\mathrm{D_{s}=8.0~kpc}$ and the limb-darkening coefficients
$\mathrm{c_{1}=0.64}$ and $\mathrm{c_{2}=0.032}$.}\label{tab1}
\end{table*}
Figure \ref{fig4a} shows a polarimetric microlensing event of a
source star with a circumstellar disk, considering three different
amounts of $\mathrm{\xi}$. Photometry and polarimetry curves are
shown in the left and right panels. The source star (grey circle),
its circumstellar disk (black ellipses) projected on the lens plane
and the lens trajectory are shown with an inset in the left-hand
panel. The simple models without disk effect are shown by red solid
lines. The thinner straight lines in the right panels represent the
polarization signal of the disk without lensing effect. The
photometric and polarimetric residuals with respect to the simple
models are plotted in the bottom panels. The polarimetric residual
is the residual in the polarization vector i.e.
$$\mathrm{|\boldsymbol{\Delta P}|=
\sqrt{P'^{2}+P^{2}-2P'P\cos2(\theta'_{p}-\theta_{p})},}$$ where the
prime symbol refers to the related quantity considering the disk
effect. The parameters used to generate these microlensing events
can be found in Table (\ref{tab1}). For the case that lens
trajectory is parallel with the semimajor axis of the disk, the
polarimetry curve has the longest time scale.

The polarization angle of an inclined disk without lensing effect is
$\mathrm{90^{\circ}}$ with respect to its semimajor axis. Whereas,
the polarization signal of the source is always normal to the
lens-source connection line (see e.g. Sajadian and Rahvar 2015).
When the lens is entering into an inclined disk normal to the disk
semimajor axis $\mathrm{(\xi=90^{\circ})}$, the polarizations of the
disk and source are normal to each other. In that case as the lens
is moving towards the source center, at some time these two
polarization signals will have the same amounts and the total
polarization signal vanishes which is shown in Figure \ref{fig4a}.
When $\mathrm{\xi=0^{\circ}}$, the polarization signals due to the
source and disk are always parallel and as a result enhances each
other. Hence, as the lens is entering the disk parallel with the
disk semimajor axis the total polarization signal is always
enhancing (see Figure \ref{fig4a}). Detecting these features in the
polarimetric curves of the microlensing events helps us to evaluate
the angle of the lens-source trajectory with respect to the disk
semimajor axis i.e. $\mathrm{\xi}$. Noting that the variation of
$\mathrm{\xi}$ has no detectable effect on the light curves and the
photometric residuals are similar for different values of
$\mathrm{\xi}$. Hence, the polarimetry observations of microlensing
events can give some information about the disk geometrical
structure.

If the impact parameter of the lens trajectory is small
$\mathrm{u_{0}\sim 0}$, the polarimetry and photometry curves almost
have symmetric shapes with respect to the time of the closest
approach. Because, the disk surface density is symmetric with
respect to the semimajor and minor axes (see Figure \ref{fig4a}).
However, due to the projection process, if $\mathrm{\varrho_{i}\cos(
i)<\rho_{\star}}$ some portion of the disk located near the inner
radius and over the y-axis is blocked by the source edge. This
causes the axial symmetry of the disk around the semimajor axis
breaks a bit and makes a small perturbation in the polarimetry and
light curves. Ignoring this perturbation due to its small amount,
the polarimetric microlensing event of the star with the
circumstellar disk while the lens impact parameter is so small
$\mathrm{(u_{0}\sim0)}$ can be degenerate by the polarimetric
microlensing event of the star with the symmetric envelop.

(b) Two microlensing events with source stars surrounded by
circumstellar disks with different parameters can have the same
polarimetry and light curves and be degenerated. In this case the
time scale of the disk crossing is a degenerate function of the disk
inclination angle $i$, the lens impact parameter $\mathrm{u_{0}}$,
the angle of the lens trajectory with respect to the semimajor axis
of the disk $\mathrm{\xi}$ and the disk equatorial optical depth
(see equation \ref{len}). Indeed, the more inclined disks have
higher intrinsic polarization signals (see Figure \ref{fig2}), but
by considering the lower optical depth and the density distribution
with the slower slope for the more inclined disk, two disks with
different inclination angles can have the same polarization signals.
Hence, the disk and lens geometrical parameters can be degenerate
with the disk surface density. Figure \ref{fig4b}, shows an example
of degenerated case for polarimetric observation with different disk
and lensing parameters. This degeneracy in the light curves of the
source stars with circumstellar disks was also mentioned by Zheng
and M\'enard (2005).

(c) The maximum disk-induced polarimetric signal happens when the
lens reaches to the disk inner radius $\mathrm{\sim\varrho_{i}}$. If
the disk inner radius is so close to the source radius, then the
maximum disk-induced polarimetric signal is almost overlapped on the
primary peaks of the polarimetric curve which occurs at
$\mathrm{u\simeq 0.96 \rho_{\star}}$ where $\mathrm{u}$ it66of the
polarimetric curve increases and the amount of this increase depends
on the disk equatorial optical depth. If the disk inner radius is
larger than the source radius, the time interval between the primary
peaks of the polarimetric curve and the peaks of the disk-induced
perturbation increases and as a result two other peaks appear in the
polarimetric curves. In Figure \ref{fig4c} we plot a polarimetric
microlensing event of a source star surrounded by a circumstellar
disk considering three different amounts of the disk inner radii.
The equatorial optical depth for three cases are aligned so that
they have the same intrinsic polarization signals. The time
intervals between the primary peaks of the polarimetric curve and
the peaks of the disk-induced perturbations, by assuming
$\mathrm{u_{0}\leq \rho_{\star}}$, are given by:
\begin{eqnarray}
\Delta t_{1,2}(t_{E})&=&\frac{\pm
2\sqrt{\varrho^{2}_{i}f^{2}-u^{2}_{0}(1+\tan^{2}i)}-u_{0}\sin
2\xi\tan^{2} i}{2f^2}
\nonumber\\&\mp&\sqrt{\rho^2_{\star}-u^2_{0}},
\end{eqnarray}
where the first term (we name it as $\mathrm{t^{d}_{1,2}(t_{E})}$)
gives the crossing times of the disk inner radius and
$\mathrm{\Delta t_{1,2}}$ are normalized to the Einstein crossing
time. By measuring these time intervals we obtain two constrains on
the geometrical parameters of the disk. If the source radius and the
lens impact parameter are inferred from the photometric
measurements, these time intervals give us a lower limit to the disk
inner radius. However, in that case for observing the disk-induced
peaks, the exposure time for polarimetric observations should be
less than $\mathrm{\Delta t_{1,2}}$. Means that we need to have at
least one data point between the primary peak and the disk-induced
peak to resolve these peaks from each other.

Note that the photometric disk-induced perturbations decrease by
increasing the disk inner radius in spite of enhancing the
equatorial optical depth (i.e. disk mass). Increasing the disk inner
radius decreases the magnification factors from the disk
contribution (as well as decreases the disk contribution in the
Stokes parameters), even though the ratio of the disk intensity to
the source intensity at the inner radius is fixed. Therefore, if we
assume that the disk inner radius is located just beyond the
condensation radius for different types of stars, the less massive
stars (late-type ones) are much suitable to be probed with the
photometric microlensing method. For these stars their condensation
radii are closer to the source centre. Cool disks which are located
beyond several Astronomical Unit (AU) from their host stars can not
most probably be detected using the gravitational microlensing.

(d) Let us assume that the lens impact parameter is in the ranges of
$\mathrm{\rho_{\star}<u_{0}\leq\varrho_{i}~f~\cos(i)}$ which means
the lens trajectory does not cross the projected source surface, but
crosses the projected disk inner radius at two times i.e.
$\mathrm{t^{d}_{1,2}(t_{E})}$. In this bypass case the primary
polarimetry curve has one peak at the time of the closest approach
(see e.g. Yushida 2006). In addition, the disk-induced polarization
signal maximizes at two moments $\mathrm{t^{d}_{1}(t_{E})}$ and
$\mathrm{t^{d}_{2}(t_{E})}$ while these times are not symmetric with
respect to the time of the closest approach i.e. $\mathrm{t_{0}}$
and the polarization signals at these two moments are not identical.
Consequently, for this event the polarimetry curve has three peaks:
two non-symmetric peaks due to the crossing the disk inner radius
and the other peak for the time of the closest approach. According
to the lens impact parameter, the disk inner radius and the disk
equatorial optical depth two disk-induced peaks can be higher or
lower than the primary peak in the polarimetry curve. In Figure
\ref{fig4d}, we plot a polarimetric microlensing event of a source
star with a disk by considering three different amounts of the disk
equatorial optical depths. Here, the lens impact parameter is
aligned so that the lens trajectory crosses the disk inner radius
but does not cross the source surface. The three mentioned peaks are
seen in this Figure. The disk-induced polarimetric peaks enlarge by
increasing the optical depth of the disk. Measuring the time
intervals between the disk-induced peaks and $\mathrm{t_{0}}$ gives
us two constraints on the disk geometrical parameters, the lens
trajectory and a lower limit on the disk inner
radius. 

(e) If the lens trajectory does not cross the disk inner radius and
the source surface i.e. $\mathrm{u_{0}>\varrho_{i}~f~\cos(i)}$, the
polarization signals due to the disk and source maximize at two
different times: the time of the closest approach of the lens from
the source center $\mathrm{t_{0}}$ and that from the disk inner
radius $\mathrm{t_{p}}$. These times do not generally coincide.
Therefore, the polarimetric curve has two peaks at these times. The
time interval between these times is a function of the geometrical
parameters of the disk and the lens trajectory which is given by:
\begin{eqnarray}
\delta
t(t_{E})=\varrho_{i}(1-b^2)\frac{\cos\xi}{\sqrt{1+b^2\cot\xi^{2}}},
\end{eqnarray}
where $\mathrm{b=\cos i}$. Measuring this time interval gives us one
constraint on the disk geometrical parameters, the lens trajectory
and a lower limit on the disk inner radius. The polarization signals
at these times depend on the disk equatorial optical depth and the
lens impact parameter. In Figure \ref{fig4e} we show a polarimetric
microlensing event of a source star surrounded by a disk so that
$\mathrm{u_{0}>\varrho_{i}~f~\cos(i)}$. There are two peaks in the
polarimetric curves corresponding to the closest approach of the
lens with respect to the disk inner radius and the source center. We
consider three different values of the impact parameter of the lens to
show its effect on the polarization signal at $\mathrm{t_{0}}$. The
residual of polarization with respect to the source star without disk
is calculated for the case of each impact parameter.

Noting that the ratio of the disk-induced peak(s) to the primary
peak in the polarimetry curves of Figures \ref{fig4d} and
\ref{fig4e} gives one constraint on the ratio of the lens impact
parameter to the disk equatorial optical depth. Comparing the
photometric and polarimetric residuals in Figures \ref{fig4a},
\ref{fig4b}, \ref{fig4c}, \ref{fig4d} and \ref{fig4e}, we notice the
maximum photometric disk-induced perturbations happen at
$\mathrm{t_{0}}$, the time of the closest approach whereas the
maximum polarimetric disk-induced perturbations occur at
$\mathrm{t^{d}_{1,2}}$ the crossing times of the disk inner radius.
Hence, although the photometric observations can be more
conservative in detecting perturbation signals due to disks, but
they can not give us any information about the disk structure
specially disk inner radius. Whereas, polarimetric observations
potentially give a lower limit to the disk inner radius.

(f) Finally, we investigate the disk-induced perturbation on the
polarization angle. Figure \ref{fig4f} shows a polarimetric
microlensing event, considering three different amounts of
$\mathrm{\xi}$. The curves of the polarization degree and its angle
are plotted in the left and right panels. The polarization vector of
the source by itself is normal to the lens-source connection line
i.e. $\mathrm{\theta_{p}=\xi+90^{\circ}}$. The existence of the disk
alters the polarization angle from this amount specially when the
lens is crossing the disk. However, by increasing the lens distance
from the source center the disk Stokes intensities decrease
significantly. Measuring the polarization angle helps to indicate
$\mathrm{\xi}$ and $i$.

According to the different panels of Figure (\ref{fig4}), it seems
that the disk-induced perturbations on the polarimetric curves of
microlensing events are larger than those on the photometric curves,
but detectability of these polarimetric signals is not necessarily
larger than that of the photometric ones. This factor depends on the
precision of the available instruments. The polarimetric precision
of the best available polarimeters can reach to $0.1$ per cent for
high amounts of the signal-to-noise ratio $\mathrm{(S/N)}$, whereas
the photometric observations of microlensing events are very
conservative. We compare photometric and polarimetric efficiencies
for detecting circumstellar disks by doing a Monte Carlo simulation
which is explained in the next section.

\section{Detectability of disk-induced perturbations}\label{Monte}
To investigate which observations of  photometry or polarimetry
is much efficient in detecting the disk-induced signatures, we
perform a Monte Carlo simulation. We first simulate an ensemble of
high-magnification and single-lens microlensing events of source
stars surrounded by circumstellar disks. By considering two useful
criteria for photometric and polarimetric observations, we
investigate if the disk-induced perturbations can be discerned in
the light and polarimetric curves. Our criterion for detectability
of disks in microlensing light curves is $\mathrm{(A'-A)/A\geq 2}$
per cent, where $A'$ and $A$ are the magnification factors with and
without disk effect.

For polarimetric observations, we assume that these observations are
done by the FOcal Reducer and low dispersion Spectrograph (FORS2)
polarimeter at Very Large Telescope (VLT) telescope. The
polarimetric precision of this polarimeter is a function of
$\mathrm{S/N}$ and improves by increasing it. The maximum
polarimetric precision which is achievable by this instrument is
$0.1$ per cent by taking one hour exposure time from a source star
brighter than $14.5$ mag \cite{Ingrosso15} which is equivalent to
$\mathrm{S/R\sim 34000}$. Our definition of $\mathrm{S/N}$ can be
found in Sajadian (2015b). We consider the following relation
between $\mathrm{S/N}$ and the polarimetric precision of the FORS2
polarimeter \cite{schemid2002}:
\begin{eqnarray}\label{sigma}
\sigma_{p}[\%]=\frac{3400}{S/R},
\end{eqnarray}
where $\mathrm{\sigma_{p}}$ is the polarimetric precision in per
cent.

For each simulated microlensing event, if
$\mathrm{u_{0}<\varrho_{i}~f~\cos(i) }$ the disk-induced polarimetric
perturbation contains two (almost similar) peaks on the both sides
of the primary peak(s) (see Figure \ref{fig4c} and \ref{fig4d}). We
investigate the detectability of the disk signal in one side of the
polarimetric curve and in the interval between the time of the
disk-induced peak $t^{d}_{1}$ and the crossing time of the disk
outer radius $\mathrm{t_{o}}$. We divide this time interval to three
parts (corresponding to three hypothetical data points) and in each
portion calculate the overall signal to noise ratio and the overall
Stokes parameters. Then, we investigate if the polarimetric
perturbation due to the disk $\mathrm{|\boldsymbol{\Delta P}|}$ is
(at least) larger than the FORS2 polarimetric precision
$\mathrm{\sigma_{p}}$ corresponding to that overall $\mathrm{S/N}$.
If for at least two consecutive data points (i.e. in two portions)
$\mathrm{|\boldsymbol{\Delta P}|> \sigma_{p}}$, the polarimetric
signal due to the disk is detectable. If $\mathrm{u_{0}>\varrho_{i}~f~
\cos(i) }$ we consider $\mathrm{2 t_{o}}$ as the time interval.

Here we use the distribution functions used to simulate microlensing
events. These functions are defining the mass of lenses, the
velocities of both sources and lenses, distribution of matter in the
Galaxy. Also we take the geometrical distributions as the source
trajectory projected on the lens plane. The generic procedure for
the Monte-Carlo Simulation is described in our previous works
\cite{sajadian12,sajadian15}.

We consider only late-type stars as microlensing sources. Because,
the condensation radii of these stars are closer to the source
centers than those for early-type stars. Hence, the peaks of the
disk-induced perturbations in the polarimetric curves occur closer
to their primary peaks. The closer distances from the source
centers, the higher magnification factors and as a result the higher
$\mathrm{S/N}$s. Indeed, the detectability of the disk-induced
polarimetric perturbations depends strongly on (i) the strength of
the perturbation peak indicated according to the disk equatorial
optical depth, a function of the disk mass and its density and (ii)
$\mathrm{S/N}$, a function of the magnification factor and the
source magnitude. On the other hand, for early-type stars the time
interval between the primary peaks in the polarimetry curves (happen
around $\mathrm{t_{\star}}$) and the disk-induced perturbation peaks
(happen near the disk inner radius) is too long, so that the
probability of detecting them decreases. Because, the polarimetry
observations of microlensing events will be most likely done when
the magnification of the source and $\mathrm{S/N}$ are enough high.
We also consider only the high-magnification and single-lens
microlensing events in which the lens impact parameter is less than
a threshold amount $\mathrm{u_{0}<u_{th}}$ and take the threshold
impact parameter $\mathrm{u_{th}=0.0008}$ which is the average
amount of $\mathrm{\rho_{\star}}$ for the Galactic microlensing
events of late-type source stars.

For the disk parameters, we assume the disk inner radius is out of
the source condensation radius and indicate this radius according to
the following analytical relation \cite{Lamers1999}:
$\mathrm{R_{h}=0.45~R_{\star}(T_{\star}(K)/1500)^{2.5}}$, where
$\mathrm{R_{h}}$ is the condensation radius, $\mathrm{T_{\star}}$ is
the effective surface temperature of the source and
$\mathrm{R_{\star}}$ is the source radius. The inclination angle of
the disk is uniformly chosen in the range of
$\mathrm{i\in[0,86^{\circ}]}$. The disk equatorial optical depth,
given in the equation (\ref{etau}), is specified according to
$\mathrm{n_{0}}$, the disk number density in the mid plane and at
the disk inner radius. $\mathrm{n_{0}}$ is in turn a function of the
disk mass $\mathrm{M_{d}}$ (equation \ref{n0}). We choose the disk
mass uniformly in the logarithmic scale over the range of
$\mathrm{M_{d}\in [10^{-7},10^{-9}]M_{\star}}$ where
$\mathrm{M_{\star}}$ is the source mass. The uniform distribution of
the disk mass in the log scale were confirmed for proto-planetary
disks \cite{protoplanet}. Indeed, we simulate only optically-thin
disks around the late-type stars located at the Galactic bulge which
have no chance to be directly detected. These Galactic bulge disks
can be detected only with lensing. The other parameters of the disk
density are chosen as follows: $\mathrm{R_{s}}$ is uniformly
selected in the range of $\mathrm{R_{s}\in [80,100]AU}$,
$\mathrm{h_{0}}$ is chosen in the range of $\mathrm{h_{0}
\in[8,12]AU}$ smoothly. $\mathrm{\gamma}$ and $\mathrm{\beta}$ are
taken uniformly in the ranges of $\mathrm{\gamma \in [1.9,2.4]}$ and
$\mathrm{\beta \in [1.2,1.3]}$.

Having performed the Monte Carlo simulation, we conclude that the
polarimetric and photometric efficiencies for detecting
optically-thin circumstellar disks around late-type source stars in
high-magnification and single-lens microlensing events are $38.6$
and $39.2$ per cent respectively. Although the disk-induced
polarimetric perturbations are larger than the photometric ones, but
the polarimetric and photometric observations have the same
efficiencies for detecting these disks.

Here, we estimate the number of detectable disks through polarimetry
and photometry observations of single-lens and high-magnification
microlensing events. The statistic of disks around the late-type
stars located at the Galactic bulge has not been studied yet. Let us
assume that the probability of having hot disks for late-type stars
located at the Galactic bulge is the same as that for nearby
late-type stars. For stars in our neighborhood, Absil et al. (2013)
studied a biased sample of $42$ nearby main-sequence stars. They
showed that $\mathrm{14\%}$ of solar-type (G and K types) stars in
their sample were associated with hot circumstellar dusts. Ertel et
al. (2014) also studied a sample of $92$ stars and obtained the
fraction $\mathrm{8\%}$ of solar-type stars showed the signatures of
hot circumstellar dusts. Accordingly, we consider the probability
that late-type stars located at the Galactic bulge have hot disks
equals to $\mathrm{f_{1}\sim 10 \%}$. About
$\mathrm{f_{2}\sim88.5\%}$ of the Galactic bulge stars are late-type
ones, i.e. K and M-type stars. The optical depth for microlensing
observations towards the Galactic bulge is $\mathrm{\tau=4.48\times
10^{-6}}$ \cite{sum05}. However, for high-magnification microlensing
events with $\mathrm{u_{0}<u_{th}}$ the optical depth reduces to
$\mathrm{\widetilde{\tau}=u_{th}^{2}~\tau}$, because the Einstein
radius for high-magnification events reduces to
$\mathrm{\widetilde{R}_{E}=R_{E}~u_{th}}$ (see e.g. Sajadian \&
Rahvar 2012). According to the definition of the optical depth, the
number of detectable disks $\mathrm{N_{d}}$ in the
high-magnification microlensing events with $\mathrm{u_{0}<u_{th}}$
is given by:
\begin{eqnarray}
N_{d}=\frac{\pi}{2}\frac{T_{obs}~N_{bg}}{\widetilde{t}_{E}}\epsilon~\widetilde{\tau}~f_{1}~f_{2},
\end{eqnarray}
where $\mathrm{\epsilon}$ is the detection efficiency, $N_{bg}$ is
the number of background stars during the observational time of
$T_{obs}$ and $\mathrm{\widetilde{t}_{E}}$ is the time scale of the
high-magnification microlensing events defined as
$\mathrm{\widetilde{t}_{E}=t_{E}~u_{th}}$. Here, we obtain the
number of detectable disks through polarimetry and photometry
observations of high-magnification microlensing events is about
$3.9$ and $4.0$ by monitoring $150$ million objects towards the
Galactic bulge during $10$ years respectively, where we set
$\mathrm{t_{E}=27}$ days \cite{OGLE3}. This number can be reliable,
as far as the assumption $\mathrm{f_{1}\sim 10 \%}$ is trusty. Note
that we simulate only optically-thin disks. The thicker disks have
more chances to be detected.
\begin{figure}
\begin{center}
\psfig{file=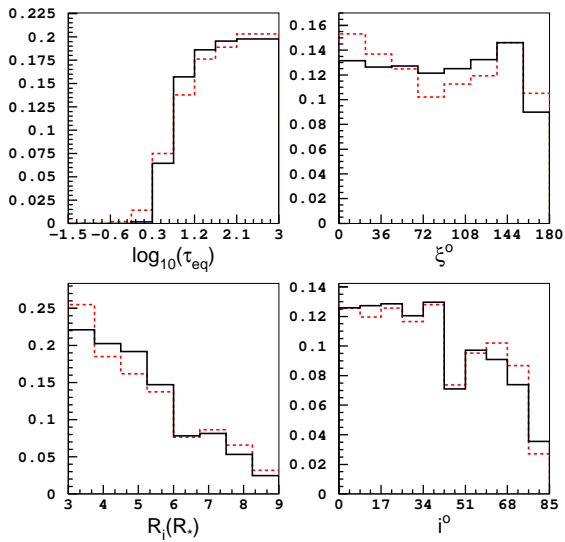,angle=0,width=8.cm,clip=} \caption{The
photometric (black solid lines) and polarimetric (red dashed lines)
efficiencies for detecting a circumstellar disk around the source
star in high-magnification microlensing events versus the disk
equatorial optical depth (top left panel), the angle of the lens
trajectory with respect to the projected semimajor axis of disk (top
right panel), the disk inner radius normalized to the source radius
(bottom left panel) and the disk inclination angle (bottom right
panel).} \label{fig5}
\end{center}
\end{figure}

In Figure (\ref{fig5}), we plot the photometric (black solid lines)
and polarimetric (red dashed lines) efficiencies for detecting
circumstellar disks around the source stars versus some relevant
parameters of the disk and the source. The detection efficiency
functions in terms of these parameters are given as follows.

(i) The first parameter is the disk equatorial optical depth,
$\mathrm{\tau_{eq}}$. The more massive disks have higher number
densities, higher equatorial optical depths and as a result the
higher Stokes intensities (see the Appendix \ref{density}). Hence,
they have higher contributions in the total Stokes parameters
(equation \ref{main}) and more chances to be detected.

(ii) The next parameter is the angle between the lens trajectory and
the semimajor axis of disk, $\mathrm{\xi^{\circ}}$. The polarimetric
detection efficiency maximizes when $\mathrm{\xi^{\circ}\sim 0,
180}$. Because, the polarimetric time scale of disk crossing
maximizes when the lens trajectory is parallel with the disk
semimajor axis (see Figure \ref{fig4a}). The photometric efficiency
does not obviously depend on $\mathrm{\xi}$.

(iii) The third parameter is the disk inner radius normalized to the
source radius, $\mathrm{R_{i}(R_{\star})}$. The maximum Stokes
intensities due to the disk occur near the disk inner radius. The
closer distance of the disk inner radius from the source center, the
higher magnification factor, the larger $\mathrm{S/N}$ and as a
result the higher probability for detecting disk-inducted
signatures.

(iv) The last parameter is the disk inclination angle
$\mathrm{i^{\circ}}$. The more inclined disks have the higher
intrinsic polarization signal and the averagely shorter time scales
of the disk crossing. Although, the first effect increases the
polarization signals but the second decreases $\mathrm{S/N}$s.
Hence, the polarimetric efficiency maximizes when $i$ is neither
$\mathrm{0^{\circ}}$ nor $\mathrm{90^{\circ}}$. By more increasing
the inclination angle two other effects appear which both decrease
efficiencies significantly: (a) some portion of the disk located
near the inner radius and above the disk semimajor axis are blocked
by the source edge when $\mathrm{\varrho_{i}\cos i<\rho_{\star}}$.
By increasing $i$, the larger portion of disk is blocked, whereas
the blocked portions have the highest Stokes intensities. (b) The
polarization of inclined disks sharply decreases by increasing the
inclination angle more than $\mathrm{80^{\circ}}$ (see Figure
\ref{fig2}).

Although the photometry and polarimetry observations of microlensing
events have the same efficiencies for detecting the signature of the
circumstellar disk around the source, but the polarimetry
observations can give some information about the disk structure, its
inner radius and its equatorial optical depth. This information can
be obtained according to the position and the strength of the
disk-induced polarimetric peaks in comparison with the primary
peaks.

On the other hand, owing to the time interval between the primary
peaks in the polarimetry curves of microlensing events and the peaks
due to the disk, detecting the signatures due to disks needs doing
polarimetry observations in the domains of the polarimetric curves,
i.e. where the lens crosses the disk inner radius. We note that the
photometric peak of disk-induced perturbations occurs at the time of
the closest approach.

\section{Conclusions}\label{result}
Gravitational microlensing is a \emph{unique} method for studying
the atmospheres of the Galactic bulge stars and their anomalies. One
of these anomalies is the existence of a circumstellar disk around
the source. In this work, we studied detecting and characterizing
optically-thin disks, contain hot dusts and are located inside one
Astronomical Unit from their host stars, around the Galactic bulge
stars through polarimetry observations of microlensing events.
Detecting these disks even around nearby stars needs high resolution
observations with interferometers \cite{Absil2013}. This kind of
disks is the most suitable candidate to be detected with lensing.
Because, the less distances from their host stars, the higher
magnification factors, the higher contributions in the total Stokes
parameters and higher $\mathrm{S/N}$s.

The sources surrounded by circumstellar disks have net polarization
signals. Lensing can magnify these polarization signals and makes
them be detected. We investigated the characteristics of the
disk-induced perturbations in polarimetric microlensing events. In
this regard, we summarized some significant points as follows:

(i) Disks cause that the time scale of the polarimetry curves in
microlensing events increases whereas they do not change the time
scale of the photometric ones. They generally break the symmetry of
the polarimetry and photometry curves around the time of the closest
approach.

(ii) There is a degeneracy between the geometrical parameters of the
disk and its projected column density in photometric or polarimetric
microlensing measurements.

(iii) If the lens moves towards the source center normal to the disk
semimajor axis in some time the total polarization signal vanishes,
whereas if the lens is interring the disk parallel with the disk
semimajor axis, the polarization signals due to the source and disk
magnify each other and the total polarization signal is always
enhancing. These points help to determine the lens trajectory with
respect to the disk semimajor axis.

(iv) The peaks of disk-induced perturbations in the light and
polarimetry curves occur at the time of the closest approach and the
time of crossing the disk inner radius respectively. According to
the lens impact parameter and the disk geometrical parameters, the
polarimetry curves of these microlensing events can have four, three
or two nonsymmetric peaks. The time intervals between these peaks
give constraints on the disk geometrical parameters and a lower
limit on the disk inner radius. The ratio of these peaks also yields
some information about the disk equatorial optical depth and the
lens impact parameter.

Finally, We compared the efficiency for detecting disks through
polarimetry observations with that through photometry observations
during microlensing events by carrying out a Monte Carlo simulation.
We concluded that the polarimetric and photometric efficiencies for
detecting an optically-thin disk around late-type sources in
high-magnification and single-lens microlensing events are similar
and are about $\mathrm{38-39}$ per cent. The number of detectable
optically-thin disks through these methods by monitoring $150$
million objects towards the Galactic bulge during $10$ years was
estimated about $\mathrm{4}$. Noting that, the polarimetry
observations help to obtain some constraints on the disk geometrical
parameters specially the disk inner radius.

\textbf{Acknowledgment} We thank R. Ignace, J. Bjorkman and Z. Zheng
and O. Absil for useful discussions and comments. We also thank the
referee for valuable comments.

\appendix
\section{Stokes intensities due to an optically-thin disk}\label{appen1}
The aim of this appendix is to describe how to
calculate the Stokes intensities of a circumstellar disk. We use the
formulation developed by Bjorkman \& Bjorkman (1994).

To describe a star with an axisymmetric circumstellar disk, two
coordinate systems are used: (a) Observer coordinate system (lens
plane) $\mathrm{(x,y,z)}$ so that the projected source center is at
the origin, the observer is on the $z$-axis at $\mathrm{+\infty}$
and the $\mathrm{z-y}$ plane contains the stellar rotation axis. (b)
Stellar coordinate system $\mathrm{(x_{\star},y_{\star},z_{\star})}$
so that $\mathrm{z_{\star}}$ is along the stellar rotation axis and
the $\mathrm{y_{\star}}$-axis is along the observer's $x$-axis. We
transform the first coordinate system to the second one by two
consecutive rotations: around $z$-axis by $\mathrm{-90^{\circ}}$ and
then around $y$-axis by the inclination angle $\mathrm{-i^{\circ}}$,
so that:
\begin{eqnarray}
x_{\star}&=&- y \cos i +  z \sin i,\nonumber\\
y_{\star}&=& x,\nonumber\\
z_{\star}&=&  y \sin i + z \cos i.
\end{eqnarray}
We use $\mathrm{(r,\theta,\phi)}$ to represent spherical stellar
coordinate, i.e.
$\mathrm{r=\sqrt{x_{\star}^{2}+y_{\star}^{2}+z_{\star}^{2}}}$,
$\mathrm{\theta=\cos^{-1}(z_{\star}/r_{\star})}$,
$\mathrm{\phi=\tan^{-1}(y_{\star}/x_{\star})}$ and
$\mathrm{(q,\alpha,z)}$ to represent cylindrical observer
coordinate, i.e. $\mathrm{q=\sqrt{x^2+y^2}}$,
$\mathrm{\alpha=\tan^{-1}(y/x)}$.

The Stokes intensities $\mathrm{\boldsymbol{I}_{\nu}}$ at frequency
$\nu$ are the solutions of the Stokes transfer equation which is
given by \cite{chandrasekhar60}:
\begin{eqnarray}
\frac{d}{dz}\boldsymbol{I}_{\nu}=-k_{\nu}\boldsymbol{I}_{\nu}+
\boldsymbol{S}_{\nu},
\end{eqnarray}
where $\mathrm{k_{\nu}}$ is the extinction coefficient which is too
small for an optically thin disk. $\mathrm{\boldsymbol{S}_{\nu}}$ is
the source function which is given by:
\begin{eqnarray}\label{sourcef}
\boldsymbol{S}_{\nu}=\frac{3 n \sigma}{4}\left(
\begin{array}{c} J+K_{zz}\\ K_{xx}-K_{yy}\\ -2K_{xy}\\0
\end{array}\right),
\end{eqnarray}
where $\sigma$ is the scattering cross section and $n$ is the number
density of disk (see the following section). $J$ and
$\mathrm{K_{ij}}$ are the zeroth and second intensity moments
respectively which are given by:
\begin{eqnarray}
J&=&\frac{1}{4\pi}\int \boldsymbol{I}_{\nu}d\Omega, \nonumber\\
K_{ij}&=&\frac{1}{4\pi} \int \boldsymbol{I}_{\nu}
\hat{n}_{i}\hat{n}_{j} d\Omega,
\end{eqnarray}
where $\mathrm{\hat{n}_{i}}$ is the unit vector in the observer
coordinate system, for $\mathrm{i\in (x,y,z)}$. Bjotkman and
Bjorkman (1994) proposed to calculate these intensity moments in the
stellar coordinate system and then using the Euler rotation matrix,
$\mathrm{\boldsymbol{R}}$, obtain the intensity moments in the
observer system. In that case, the convenient Euler angles are
$\mathrm{(\theta,\phi,i)}$ so that
$\mathrm{\boldsymbol{R}=\boldsymbol{R}_{r}(\pi/2)\boldsymbol{R}_{\phi}(i)\boldsymbol{R}_{r}(-\phi)\boldsymbol{R}_{\phi}(-\theta)}$.
This matrix relates the intensity moments in the spherical polar
stellar coordinate $\mathrm{(r,\theta,\phi)}$ to the cartesian
observer system. Its rows 1-3 correspond to $z$, $x$ and $y$ and
columns 1-3 correspond to $r$, $\mathrm{\theta}$ and
$\mathrm{\phi}$. In that case the intensity moments are transformed
as follows:
\begin{eqnarray}
J&=&J',\nonumber\\
\boldsymbol{K}&=&\boldsymbol{R}\boldsymbol{K}'\boldsymbol{R}^{T},
\end{eqnarray}
where $\mathrm{J'}$ and $\mathrm{\boldsymbol{K}'}$ are the related
intensity moments in the spherical stellar coordinate system which
are given by:
\begin{eqnarray}
J'&=&\frac{1}{4\pi}\int_{\Omega^{\star}}I^{\star}d\Omega', \nonumber\\
K'_{ij}&=&\frac{1}{4\pi}\int_{\Omega^{\star}}I^{\star}\hat{n}'_{i}\hat{n}'_{j}d\Omega',
\end{eqnarray}
where $\mathrm{\Omega^{\star}}$ is the solid angle subtended by the
star.
The unit vectors $\mathrm{\hat{n}'_{i}}$ are given by:
\begin{eqnarray}
\hat{n}'_{r}&=&\cos \theta', \nonumber\\
\hat{n}'_{\theta}&=&\sin\theta'\cos\phi',\nonumber\\
\hat{n}'_{\phi}&=&\sin\theta'\sin\phi',
\end{eqnarray}
and $\mathrm{d\Omega'=\sin\theta'd\theta'd\phi'}$.
$\mathrm{I^{\star}}$ is the unpolarized stellar intensity at the
stellar surface at the frequency $\mathrm{\nu}$ which has a constant
amount. Indeed we assume that the disk and stellar atmosphere both
are optically thin.

The emergent Stokes intensities are obtained by integrating over the
source function:
\begin{eqnarray}\label{App16}
\boldsymbol{I}(x,y)=\int_{z_{min}}^{+\infty}e^{-\tau(z)}\boldsymbol{S}_{\nu}dz
\end{eqnarray}
where $\mathrm{z_{min}}$ is the location where the line of sight
intersects the stellar surface which is given by:
\begin{eqnarray}
z_{min}(q)=\begin{array}{c}\sqrt{R_{\star}^{2}-q^{2}}~~~~~~~~(q<R_{\star})\\-\infty~~~~~~~~~~~~~~~~~~(q\geq
R_{\star})\end{array}
\end{eqnarray}
where $\mathrm{R_{\star}}$ is the source radius. The optical depth
is $\mathrm{\tau(z)=\int_{z}^{+\infty}k_{\nu}~dz}$. Note that we
separately calculate the contribution of the source Stokes
intensities which is due to the stellar atmosphere (see equation
\ref{main}).

\section{Disk density}\label{density}
We use the following model for the number density in the disk
\cite{HST1996}:
\begin{eqnarray}
n(R,z)=n_{0} (\frac{R_{c}}{R})^{\gamma}\exp[-\frac{z^2}{2H(R)^{2}}],
\end{eqnarray}
where $\mathrm{R_{c}}$ is the length scale, $\mathrm{\gamma}$
specifies the radial dependence of the disk surface density and
$\mathrm{H(R)=h_{0}(\frac{R}{R_{s}})^{\beta}}$ is the disk thickness
at the redial distance of $R$ from the disk center. $\mathrm{h_{0}}$
is the hight scale at the radial distance $\mathrm{R_{s}}$.
$\mathrm{n_{0}}$ is the number density in the mid-plane and at the
radial distance $\mathrm{R_{c}}$ which is a function of the disk
mass $\mathrm{M_{d}}$ \cite{protoplanet}:
\begin{eqnarray}\label{n0}
n_{0}=\frac{(2-\gamma+\beta)}{(2\pi)^{3/2}h_{0}R^{2}_{i}m_{0}}(\frac{R_{s}}{R_{i}})^{\beta}M_{d},
\end{eqnarray}
where $\mathrm{m_{0}\simeq 2.3 m_{p}}$ is the mean molecular weight
and $\mathrm{m_{p}}$ is the proton mass \cite{protoplanet2}. For
this disk density, the equatorial optical depth is given by:
\begin{eqnarray}\label{etau}
\tau_{eq}=\int_{R_{i}}^{\infty}dR n(R,0) \sigma=n_{0}\sigma
R_{i}(\frac{R_{c}}{R_{i}})^{\gamma}/(\gamma-1).
\end{eqnarray}
Throughout the paper, we set $\mathrm{R_{c}=R_{i}}$. Hence the disk
equatorial optical depth is $\mathrm{\tau_{eq}=n_{0}\sigma
R_{i}/(\gamma-1)}$. The source function, given by equation
(\ref{sourcef}), is proportional to
$\mathrm{\tau_{eq}(\gamma-1)/R_{i}}$. Therefore, the disk Stokes
intensities and as a result the disk polarization signal are
proportional to the disk equatorial optical depth.
\end{document}